\documentclass[preprint,aps,prd,nofootinbib,showpacs,superscriptaddress,11pt]{revtex4}
\usepackage{txfonts}
\usepackage[colorlinks, linkcolor=blue, citecolor=blue, urlcolor=blue]{hyperref}
\usepackage{float}
\usepackage{graphicx}
\usepackage{dcolumn}
\usepackage{bm}
\usepackage{amssymb}
\usepackage{latexsym}

\newcommand{\be}{\begin{equation}}
\newcommand{\ee}{\end{equation}}
\newcommand{\bq}{\begin{eqnarray}}
\newcommand{\eq}{\end{eqnarray}}

\begin{document}

\preprint{USTC-ICTS-12-07}

\title{Revisit of the Interaction between Holographic Dark Energy and Dark Matter}

\author{Zhenhui Zhang}
\email{zhangzhh@mail.ustc.edu.cn}
\affiliation{Department of Modern Physics, University of Science and Technology of China, Hefei 230026, China}
\affiliation{Institute of Theoretical Physics, Chinese Academy of Sciences, Beijing 100190, China}

\author{Song Li}
\email{sli@itp.ac.cn}
\affiliation{Institute of Theoretical Physics, Chinese Academy of Sciences, Beijing 100190, China}
\affiliation{Kavli Institute for Theoretical Physics China, Chinese Academy of Sciences, Beijing 100190, China}
\affiliation{Key Laboratory of Frontiers in Theoretical Physics, Chinese Academy of Sciences, Beijing 100190, China}

\author{Xiao-Dong Li}
\email{renzhe@mail.ustc.edu.cn}
\affiliation{Department of Modern Physics, University of Science and Technology of China, Hefei 230026, China}
\affiliation{Institute of Theoretical Physics, Chinese Academy of Sciences, Beijing 100190, China}
\affiliation{Kavli Institute for Theoretical Physics China, Chinese Academy of Sciences, Beijing 100190, China}
\affiliation{Interdisciplinary Center for Theoretical Study, University of Science and Technology of China, Hefei 230026, China}

\author{Xin Zhang}
\email{zhangxin@mail.neu.edu.cn} \affiliation{Department of Physics,
College of Sciences, Northeastern University, Shenyang 110004,
China} \affiliation{Center for High Energy Physics, Peking
University, Beijing 100080, China}

\author{Miao Li}
\email{mli@itp.ac.cn}
\affiliation{Institute of Theoretical Physics, Chinese Academy of Sciences, Beijing 100190, China}
\affiliation{Kavli Institute for Theoretical Physics China, Chinese Academy of Sciences, Beijing 100190, China}
\affiliation{Key Laboratory of Frontiers in Theoretical Physics, Chinese Academy of Sciences, Beijing 100190, China}

\begin{abstract}

In this paper we investigate the possible direct, non-gravitational
interaction between holographic dark energy (HDE) and dark matter.
Firstly, we start with two simple models with the interaction terms
$Q \propto \rho_{dm}$ and $Q \propto \rho_{de}$, and then we move on
to the general form $Q \propto \rho_m^\alpha\rho_{de}^\beta$. The
cosmological constraints of the models are obtained from the joint
analysis of the present Union2.1+BAO+CMB+$H_0$ data. We find that
the data slightly favor an energy flow from dark matter to dark
energy, although the original HDE model still lies in the 95.4\%
confidence level (CL) region. For all models we find $c<1$ at the
95.4\% CL. We show that compared with the cosmic expansion, the
effect of interaction on the evolution of $\rho_{dm}$ and
$\rho_{de}$ is smaller, and the relative increment (decrement)
amount of the energy in the dark matter component is constrained to
be less than 9\% (15\%) at the 95.4\% CL. By introducing the
interaction, we find that even when $c<1$ the big rip still can be
avoided due to the existence of a de Sitter solution at
$z\rightarrow-1$. We show that this solution can not be accomplished
in the two simple models, while for the general model such a
solution can be achieved with a large $\beta$, and the big rip may
be avoided at the 95.4\% CL.

\end{abstract}


\maketitle

\section{Introduction}

Cosmological observations such as the type Ia supernovae (SNIa)
\cite{Riess}, the cosmic microwave background (CMB) \cite{spergel}
and the large scale structure (LSS) \cite{Tegmark} all indicate that
the universe is undergoing an accelerating expansion. This implies
the existence of a mysterious component, named dark energy, which
has negative pressure and takes the largest proportion of the total
density in the present universe. In the last decade, lots of efforts
have been made to understand  dark energy \cite{DEReview}, yet we
still know little about its nature.

In this paper we discuss the possible direct, non-gravitational
interaction between the dark sectors in the framework of the
holographic dark energy model, which is a quantum gravity approach
to the dark energy problem \cite{Witten:2000zk,Holography,cohen99}.
In this model, the vacuum energy is viewed as dark energy, and is
related to the event horizon of the universe when we require that
the zero-point energy of the system should not exceed the mass of a
black hole with the same size. In this way, we have the holographic
dark energy (HDE) density~\cite{li04}
\begin{equation}\label{eq:rhoHDE}
\rho_{de} = 3c^2M^2_{Pl}R_h^{-2},
\end{equation}
where $c$ is a dimensionless model parameter, which can only be
determined by observations,
$M_{Pl}$ is the reduced Planck mass,
and $R_h$ is the future event horizon of
the universe, defined as
\begin{equation}\label{eq:rh}
R_h=a\int^\infty_t{dt \over a}=a\int^\infty_a{da\over Ha^2}.
\end{equation}
The HDE model has been proved to be a competitive and promising dark
energy candidate. It can theoretically explain the coincidence
problem \cite{li04}, and is proven to be perturbational stable
\cite{HDEstable}. It is also favored by the observational data
\cite{holoobs09}. For more studies on the HDE model, see, e.g.,
\cite{casimirHDE,refHDE,holode1,holode2}.

The HDE model with some interaction between dark energy and dark
matter (hereafter IHDE model) was firstly studied by Wang {\it et
al.} in \cite{IntBinWang}. If dark energy interacts with cold dark
matter,the continuity equations for them are
$\dot\rho_{de}+3H(\rho_{de}+p_{de})=-Q$ and
$\dot\rho_{dm}+3H\rho_{dm}=Q, $ where $Q$ phenomenologically
describes the interaction.
The interaction between the dark sectors in the HDE model has been
extensively studied in, e.g., \cite{intholo}. It was found that the
introduction of interaction may not only alleviate the cosmic
coincidence problem, but also help to avoid the future big-rip
singularity.

There are various choices for the forms of $Q$. The most common
choice is
\begin{equation}\label{Q1}
Q = \Gamma H \rho,
\end{equation}
where $\Gamma$ is a dimensionless constant, and $\rho$ is taken to
be the density of dark energy, dark matter, or the sum of them.
These models are mathematically simple, so they are useful for
phenomenology. However, it is difficult to see how they can emerge
from a physical description of dark sector interaction. It is
expected that the interaction is determined by the local properties
of the dark sectors, i.e., $\rho_{dm}$ and $\rho_{de}$, but it is
hard to understand why the interaction term must be proportional to
the Hubble expansion rate $H$. A more natural form of the
interaction was proposed by Ma {\it et al.} \cite{YZMa}, where the
interaction term takes the form
\begin{equation}\label{Q2}
Q \propto \rho_{dm}^\alpha \rho_{de}^\beta.
\end{equation}
In this description the interaction term only depends on the local
energy densities of the dark sectors, and is thus more physically
plausible. When $\alpha=1$ and $\beta=0$, the interaction term
$Q\propto \rho_{dm}$ is the exact form we expected in the case of
the dark matter decay. Similarly, the case of $\alpha=0$ and
$\beta=1$ corresponds to the dark energy decay. Moreover, in more
complicated forms of interaction, like scattering, one may expect
the existence of both $\rho_{dm}$ and $\rho_{de}$. Thus, Eq.
(\ref{Q2}) seems a more natural and physically plausible form to
describe the interaction.

In this paper we revisit the interaction between the HDE and dark
matter by considering the interaction with the form $Q\propto
\rho_{dm}^\alpha \rho_{de}^\beta$. To perform an overall analysis,
we will take into consideration as many factors as possible.
Different from many works in which the contribution of baryons is
ignored, we will consider both baryon and dark matter components in
the interacting models. Moreover, we will also discuss the models in a
non-flat universe. As argued in Ref.~\cite{Clarkson:2007bc}, the
studies of dark energy, and in particular, of observational data,
should include $\Omega_{k0}$ as a free parameter to be fitted
alongside the $w(z)$ parameters. Another reason why we take the
spatial curvature into consideration is the possible correlation
between the curvature and the interaction. As pointed out in
Ref.~\cite{KIHDE}, compared with the flat universe, a much stronger
interaction can be allowed by the data in a non-flat universe.

There are many interesting issues worth investigating in the IHDE
models. For example, the properties of the interaction term, the
direction of the energy flow, the evolution of the dark matter and
dark energy densities, the fate of the universe, and so on. In the
IHDE model with $Q\propto \rho_{dm}^\alpha \rho_{de}^\beta$, some of
these issues have been discussed in \cite{YZMa}. In this paper, we
will perform a more comprehensive exploration of these issues.

This paper is organized as follows. In Sec. II, we derive the basic
equations for the IHDE models. In Sec. III, we introduce the
methodology and data used in this work. In Sec. IV, we discuss the
cosmological interpretations of the three IHDE models, including two
simple models with $Q\propto \rho_{dm}$, $Q\propto \rho_{de}$ and
one general model with $\alpha$ and $\beta$ treated as free
parameters. We firstly study the cosmological constraints on these
models, and then discuss the fate of the universe in the models. At
last, we give some concluding remarks in Sec. V. In this work, we
assume today's scale factor $a_0 = 1$, so the redshift $z$ satisfies
$z = 1/a - 1$; the subscript ``0'' always indicates the present
value of the corresponding quantity, and the unit with $c = \hbar =
1$ is used.

\section{Interacting Holographic Dark Energy Model in a Non-flat Universe}

In this section, we give the basic equations for the IHDE model in a non-flat universe.

\subsection{Friedmann equations in a non-flat universe}

In a spatially non-flat Friedmann-Robertson-Walker universe, the Friedmann equation can be
written as
\begin{equation}
\label{eq:FE1} 3M_{Pl}^2 H^2=\rho_{dm}+\rho_b+\rho_r+\rho_k+\rho_{de},
\end{equation}
where $\rho_k=-3M_{Pl}^2\frac{k}{a^2}$ is the effective energy density of
the curvature component. For convenience, we define the fractional
energy densities of the various components, i.e.,
\begin{equation}
\label{eq:DefOmega_k} \Omega_k={-k \over
H^2a^2}={\rho_k \over \rho_c},\ \ \ \ \
\Omega_{de}={\rho_{de} \over \rho_c},\ \ \ \ \
\Omega_{dm}={\rho_{dm} \over \rho_c},\ \ \ \
\Omega_{b}={\rho_{b} \over \rho_c},\ \ \ \
\Omega_r ={\rho_r \over  \rho_c},
\end{equation}
where $\rho_c=3M_{Pl}^2H^2$ is the critical density of the universe.
The subscripts, $k$, $de$, $dm$, $b$ and $r$, represent curvature,
dark energy, dark matter, baryon and radiation, respectively. By
definition, we have
\begin{equation}
\label{eq:AllOmega} \Omega_{de}+\Omega_{dm}+\Omega_b+\Omega_r+\Omega_k=1.
\end{equation}

With the existence of interaction between the dark sectors, the
energy conservation equations for the components in the universe
take the forms
\begin{equation}
\label{eq:IHDEeq0}\dot\rho_{dm}+3H\rho_{dm}=Q,
\end{equation}
\begin{equation}
\label{eq:IHDEeq1}\dot\rho_{de}+3H(\rho_{de}+p_{de})=-Q,
\end{equation}
\begin{equation}
\label{eq:IHDEeq2}\dot\rho_b+3H\rho_b=0,
\end{equation}
\begin{equation}
\label{eq:IHDEeq3}\dot\rho_r+4H\rho_r=0,
\end{equation}
\begin{equation}
\label{eq:IHDEeq4}\dot\rho_k+2H\rho_k=0.
\end{equation}
As mentioned above, $Q$ denotes the phenomenological interaction
term. Combining Eqs.~(\ref{eq:IHDEeq0})--(\ref{eq:IHDEeq4})
together, we can obtain the form of $p_{de}$,
\begin{equation}\label{eq:pde}
p_{de}=-\frac{2}{3}\frac{\dot H}{H^2}\rho_c-\rho_c-{1\over3}\rho_r+{1\over3}\rho_k.\label{eq:pde}
\end{equation}
Substituting $p_{de}$ into Eq. (\ref{eq:IHDEeq1}), we have
\begin{equation}
\Big(2{\dot H\over H}+{\dot\Omega_{de}\over\Omega_{de}}+3H
\Big)\rho_{de}+H(\rho_k-\rho_r)-\Big(2{\dot H\over H}+3H\Big)\rho_c
=-Q.
\end{equation}
Dividing the above equation by $\rho_c$,
we get a derivative equation of $\dot H$ and $\dot \Omega_{de}$,
\begin{equation}
\label{eq:OH2} 2(\Omega_{de}-1){\dot H\over
H}+\dot\Omega_{de}+H(3\Omega_{de}-3+\Omega_k-\Omega_r)=-H\Omega_I,
\end{equation}
where we have defined the effective dimensionless quantity for
interaction, $\Omega_I$, with the form
\begin{equation}
\Omega_I\equiv\frac{Q}{H(z)\rho_c}.
\end{equation}

\subsection{HDE in a non-flat universe}

From the energy density of the HDE, Eq.~(\ref{eq:rhoHDE}), we have
\begin{equation}
\label{ea:L0} L={c \over H\sqrt{\Omega_{de}}}.
\end{equation}
Following Ref.~\cite{holode1}, in a non-flat universe, the IR
cut-off length scale $L$ takes the form
\begin{equation}
\label{eq:L1} L=ar(t),
\end{equation}
and $r(t)$ satisfies
\begin{equation}
\label{eq:r(t)1}  \int_0^{r(t)} {dr \over
\sqrt{1-kr^2}}=\int_t^{+\infty}{dt\over a(t)}.
\end{equation}
By carrying out the integration, we have
\begin{equation}
\label{eq:r(t)2} r(t)={1\over\sqrt{k}}\sin
\Big(\sqrt{k}\int_t^{+\infty} {dt \over a}\Big)={1\over\sqrt{k}}\sin
\Big(\sqrt{k}\int_{a(t)}^{+\infty} {da \over {Ha^2}}\Big).
\end{equation}
Equation (\ref{eq:L1}) leads to another equation about $r(t)$,
namely,
\begin{equation}
\label{eq:r(t)3.0} r(t)={L\over
a}={c\over\sqrt{\Omega_{de}}Ha}.
\end{equation}
Combining Eqs. (\ref{eq:r(t)2}) and (\ref{eq:r(t)3.0}) yields
\begin{equation}
\label{eq:OL1.1}\sqrt{k}\int_t^{+\infty}{dt\over
a}=\arcsin{c\sqrt{k}\over \sqrt{\Omega_{de}}aH}.
\end{equation}
Taking derivative of Eq.~(\ref{eq:OL1.1}) with respect to $t$, one
can get
\begin{equation}
\label{eq:OL1.3}
{\dot\Omega_{de}\over2\Omega_{de}}+H+{\dot H\over H}=\sqrt{{\Omega_{de}H^2\over c^2}-{k\over a^2}}.
\end{equation}

\subsection{Evolution equations of $E(z)$ and $\Omega_{de}(z)$ in an IHDE scenario}

Combining Eq.~(\ref{eq:OH2}) with Eq.~(\ref{eq:OL1.3}), we
eventually obtain the following two equations governing the
dynamical evolution of the IHDE model in a non-flat universe,
\begin{equation}
\label{eq:OH3}{1\over E(z)}{dE(z) \over dz}
=-{\Omega_{de}\over
1+z}\left({\Omega_k-\Omega_r-3+\Omega_I\over2\Omega_{de}}+{1\over2}+\sqrt{{\Omega_{de}\over
c^2}+\Omega_k} \right),
\end{equation}
\begin{equation}
\label{eq:OH4}
{d\Omega_{de}\over dz}=
-{2\Omega_{de}(1-\Omega_{de})\over 1+z}\left(\sqrt{{\Omega_{de}\over
c^2}+\Omega_k}+{1\over2}-{\Omega_k-\Omega_r+\Omega_I\over 2(1-\Omega_{de})}\right),
\end{equation}
where $E(z)\equiv H(z)/H_0$ is the dimensionless Hubble expansion
rate. Equations (\ref{eq:OH3}) and (\ref{eq:OH4}) can be solved
numerically and will be used in the data analysis procedure. Notice
that we have
\begin{equation}
\Omega_k(z)=\frac{\Omega_{k0}(1+z)^2}{E(z)^2},\ \  \Omega_r(z)=\frac{\Omega_{r0}(1+z)^4}{E(z)^2},\ \
\Omega_b(z)=\frac{\Omega_{b0}(1+z)^3}{E(z)^2},
\end{equation}
and the fractional density of dark matter is given by
$\Omega_{dm}(z)=1-\Omega_k(z)-\Omega_{de}(z)-\Omega_r(z)-\Omega_b(z)$.
The values of $\Omega_{b0}$ and $\Omega_{r0}$, for simplicity, are
determined from the 7-yr WMAP observations~\cite{WMAP7},
\begin{equation}
  \Omega_{b0} = 0.02253h^{-2},
\end{equation}
\begin{equation}
\Omega_{r0} = \Omega_{\gamma 0}(1+0.2271 N_{eff}),\ \ \ \Omega_{\gamma 0} = 2.469 \times 10^{-5} h^{-2},\ \ \  N_{eff}=3.04,
\end{equation}
where $\gamma$ represents photons, and $N_{eff}$ is the
effective number of neutrino species.

\subsection{Models}

In this paper we consider the interaction term with the form $Q
\propto \rho_{dm}^\alpha\rho_{de}^\beta$. For concreteness, we
express the interaction term as
\begin{equation}\label{eq:Q}
Q = \Gamma H_0 \ \frac{\rho_{dm}^\alpha\ \rho_{de}^\beta} {\ \rho_{c0}^{\alpha+\beta-1}},
\end{equation}
which can also be expressed as $Q = \Gamma H_0 \rho_{c0}
E(z)^{2(\alpha+\beta)} \Omega_{dm}(z)^\alpha \Omega_{de}(z)^\beta$,
and the parameter $\Gamma$ is dimensionless.

We will investigate three IHDE models in this paper.
Firstly, we discuss two simple models with fixed $\alpha$ and $\beta$.
The first model (hereafter IHDE1) is the case of $\alpha=1$ and $\beta=0$,
and thus
\begin{equation}
Q= \Gamma H_0 \rho_{dm},\ \ \  \Omega_I = \Gamma \Omega_{dm}(z)/E(z).
\end{equation}
When $\Gamma>0$, the energy transfer corresponds to the decay of
dark energy into dark matter; {\it vice versa}. This model is
similar to the model proposed in \cite{RoyMaartens}, where the same
form is introduced in the interaction between dark matter and
quintessence.

The second model (hereafter IHDE2) considered in this paper is the case of $\alpha=0$ and $\beta=1$.
Correspondingly, we have
\begin{equation}
Q= \Gamma H_0 \rho_{de},\ \ \  \Omega_I = \Gamma \Omega_{de}(z)/E(z).
\end{equation}

Finally, we also consider a more general model (here after IHDE3)
where $\alpha$ and $\beta$ are treated as free parameters. The
formula of $Q$ has been given in Eq. (\ref{eq:Q}), while the
dimensionless quantity for interaction defined previously is
\begin{equation}
\Omega_I = \Gamma E(z)^{2(\alpha+\beta)-3} \Omega_{dm}(z)^\alpha \Omega_{de}(z)^\beta.
\end{equation}

\section{Methodology}

For the IHDE models in a non-flat universe, there are six free
parameters: $c$, $\Omega_{dm0}$, $\Omega_{k0}$, $\Gamma$, $\alpha$
and $\beta$. We will constrain them by using the latest observational data.
As a comparison, the $\Lambda$CDM model and the HDE model with spatial curvature but
without interaction (namely, $\Omega_{k0}\neq 0$ but $Q=0$) will
also be investigated.

In this work, we adopt the $\chi^2$ statistic to estimate the model
parameters. For a physical quantity $\xi$ with experimentally
measured value $\xi_{obs}$, standard deviation $\sigma_{\xi}$ and
theoretically predicted value $\xi_{th}$, the $\chi^2$ function
takes the form
\begin{equation}
\chi^2_{\xi}={(\xi_{obs}-\xi_{th})^2\over \sigma^2_{\xi}}~.
\end{equation}
The total $\chi^2$ is the sum of all $\chi^2_{\xi}$s, i.e.
\begin{equation}
\chi^2=\sum_{\xi}\chi^2_{\xi}~.
\end{equation}
One can determine the best-fit model parameters by minimizing the total $\chi^2$.
Moreover, by calculating $\Delta \chi^2 \equiv \chi^2-\chi^2_{\rm min}$,
one can determine the 68.3\% and the 95.4\% CL ranges of a specific model.

In this work, we determine the best-fit values and the 68.3\%
and 95.4\% CL ranges of the model parameters by using the Markov Chain Monte Carlo (MCMC)
technique. We modify the publicly available CosmoMC package
\cite{COSMOMC} and generate $10^6$--$10^7$ samples for each set of
results presented in this paper.

For data, we use the Union2.1 SNIa sample \cite{Union2.1}, the CMB
anisotropy data from the 7-yr WMAP observations \cite{WMAP7}, the
BAO results from the SDSS DR7 \cite{SDSSDR7}, 6dFGS \cite{6dFGS} and
WiggleZ Dark Energy Survey \cite{WiggleZ}, and the Hubble constant
measurement from the WFC3 on the HST \cite{HSTWFC3}. In the
following, we briefly describe how these data are included into the
$\chi^2$ analysis.

\subsection{The SNIa data}

First we start with the SNIa observations. We use the latest
Union2.1 sample including 580 SNIa that are given in terms of the
distance modulus $\mu_{obs}(z_i)$ \cite{Union2.1}. The theoretical
distance modulus is defined as
\begin{equation}
\mu_{th}(z_i)\equiv 5 \log_{10} {D_L(z_i)} +\mu_0,
\end{equation}
where $\mu_0\equiv 42.38-5\log_{10}h$ with $h$ the Hubble constant
$H_0$ in units of 100 km/s/Mpc, and the Hubble-free luminosity
distance $D_L=H_0d_L$ is
\begin{equation}
D_L(z)={1+z\over \sqrt{|\Omega_{k0}|}}\textrm{sinn}\Big(
\sqrt{|\Omega_{k0}|}\int_0^z{dz'\over E(z')} \Big),
\end{equation}
where
\begin{displaymath}
\textrm{sinn}\left(x\right) = \left\{
\begin{array}{ll}
\textrm{sin}(x), & \textrm{if $\Omega_{k0}<0$},\\
x, & \textrm{if $\Omega_{k0}=0$},\\
{\textrm{sinh}(x)}, &
\textrm{if $\Omega_{k0}>0$}.
\end{array} \right.
\end{displaymath}

The $\chi^2$ function for the SNIa data is
\begin{equation}
\chi^2_{SN}=\sum\limits_{i=1}^{580}{[\mu_{obs}(z_i)-\mu_{th}(z_i)]^2\over
\sigma_i^2},\label{ochisn}
\end{equation}
where $\mu_{obs}(z_i)$ and $\sigma_i$ are the observed value and the
corresponding 68.3\% error of distance modulus for each supernova,
respectively. For convenience, people often analytically marginalize
the nuisance parameter $\mu_0$ (i.e., the Hubble constant $H_0$)
when calculating $\chi^2_{SN}$ \cite{Perivolaropoulos}.

It should be stressed that Eq.~(\ref{ochisn}) only considers the
statistical errors from SNIa, and ignores the systematic errors from
SNIa. To include the effect of systematic errors into our analysis,
we will follow the prescription for using the Union2.1 compilation
provided in \cite{Union Web}. The key of this prescription is a $580
\times 580$ covariance matrix, $C_{SN}$, which captures the
systematic errors from SNIa (This covariance matrix with systematics
can be downloaded from \cite{Union Web}). Utilizing $C_{SN}$, we can
calculate the following quantities
\begin{equation}
A=(\mu^{obs}_i-\mu^{th}_i)(C_{SN}^{-1})_{ij}(\mu^{obs}_j-\mu^{th}_j),\label{chisna}
\end{equation}
\begin{equation}
B=\sum\limits_{i=1}^{580}{(C_{SN}^{-1})_{ij}(\mu^{obs}_j-\mu^{th}_j)},\label{chisnb}
\end{equation}
\begin{equation}
C=\sum\limits_{i,j=1}^{580}{(C_{SN}^{-1})_{ij}}\label{chisnc},
\end{equation}
and the $\chi^2$ function for the SNIa data is \cite{Union Web}
\begin{equation}
\chi^2_{SN}=A-\frac{B^{2}}{C}.\label{chisn}
\end{equation}
Different from Eq.~(\ref{ochisn}), this formula includes the effect
of systematic errors from SNIa.

\subsection{The CMB data}

Here we use the ``WMAP distance priors'' given by the 7-yr WMAP
observations \cite{WMAP7}. The distance priors include the
``acoustic scale'' $l_A$, the ``shift parameter'' $R$, and the
redshift of the decoupling epoch of photons $z_*$. The acoustic
scale $l_A$, which represents the CMB multipole corresponding to the
location of the acoustic peak, is defined as \cite{WMAP7}
\begin{equation}
\label{ladefeq} l_A\equiv (1+z_*){\pi D_A(z_*)\over r_s(z_*)}~.
\end{equation}
Here $D_A(z)$ is the proper angular diameter distance, given by
\begin{equation}
D_A(z)=d_L(z)/(1+z)^2,
\label{eq:da}
\end{equation}
and $r_s(z)$ is the comoving sound horizon size, given by
\begin{equation}
r_s(z)=\frac{1} {\sqrt{3}}  \int_0^{1/(1+z)}  \frac{ da } { a^2H(a)
\sqrt{1+(3\Omega_{b0}/4\Omega_{\gamma0})a} }~,
\label{eq:rs}
\end{equation}
where $\Omega_{b0}$ and $\Omega_{\gamma0}$ are the present baryon
and photon density parameters, respectively. As mentioned above, we
adopt the best-fit values, $\Omega_{b0}=0.02253 h^{-2}$ and
$\Omega_{\gamma0}=2.469\times10^{-5}h^{-2}$ (for $T_{cmb}=2.725$ K),
given by the 7-yr WMAP observations \cite{WMAP7}. The fitting
function of $z_*$ was proposed by Hu and Sugiyama~\cite{Hu:1995en}:
\begin{equation}
\label{zstareq} z_*=1048[1+0.00124(\Omega_{b0}
h^2)^{-0.738}][1+g_1(\Omega_{m0} h^2)^{g_2}]~,
\end{equation}
where
\begin{equation}
g_1=\frac{0.0783(\Omega_{b0} h^2)^{-0.238}}{1+39.5(\Omega_{b0} h^2)^{0.763}}~,
\quad g_2=\frac{0.560}{1+21.1(\Omega_{b0} h^2)^{1.81}}~.
\end{equation}
Here the subscript $m$ denotes the matter component, i.e.,
$\Omega_m=\Omega_{dm}+\Omega_b$. In addition, the shift parameter
$R$ is defined as \cite{Bond97}
\begin{equation}
\label{shift} R(z_*)\equiv \sqrt{\Omega_{m0} H_0^2}(1+z_*)D_A(z_*)~.
\end{equation}
This parameter has been widely used to constrain various cosmological models \cite{Add3}.

As shown in \cite{WMAP7}, the $\chi^2$ function of the CMB data is
\begin{equation}
\chi_{CMB}^2=(x^{obs}_i-x^{th}_i)(C_{CMB}^{-1})_{ij}(x^{obs}_j-x^{th}_j),\label{chicmb}~
\end{equation}
where $x_i=(l_A, R, z_*)$ is a vector, and $(C_{CMB}^{-1})_{ij}$ is
the inverse covariance matrix. The 7-yr WMAP observations
\cite{WMAP7} have provided the maximum likelihood values:
$l_A(z_*)=302.09$, $R(z_*)=1.725$, and $z_*=1091.3$. The inverse
covariance matrix was also given in \cite{WMAP7},
\begin{equation}
(C_{CMB}^{-1})=\left(
  \begin{array}{ccc}
    2.305 & 29.698 & -1.333 \\
    29.698 & 6825.27 & -113.180 \\
    -1.333 & -113.180  &  3.414 \\
  \end{array}
\right).
\end{equation}

\subsection{The BAO data}

In this paper we use the BAO data from the SDSS DR7 \cite{SDSSDR7}, the 6dFGS \cite{6dFGS}, and the WiggleZ Dark Energy Survey \cite{WiggleZ}.
In the following, we will describe the BAO distance measurements of these projects,
and introduce how to add them into the $\chi^2$ statistics.

One effective distance measure is $D_V(z)$, which can be obtained
from the spherical average \cite{Eisenstein}
\begin{equation}
 D_V(z) \equiv \left[(1+z)^2D_A^2(z)\frac{z}{H(z)}\right]^{1/3},
\end{equation}
where $D_A(z)$ is the proper angular diameter distance.

For the 6dFGS and SDSS DR7 data, we use the quantity $d_{z}\equiv
r_s(z_d)/D_V(z)$. The expression of $r_s$ is given in
Eq.(\ref{eq:rs}), and $z_d$ denotes the redshift of the drag epoch,
whose fitting formula is proposed by Eisenstein and
Hu~\cite{BAODefzd},
\begin{equation}
\label{Defzd} z_d={1291(\Omega_{m0}h^2)^{0.251}\over 1+0.659(\Omega_{m0}h^2)^{0.828}}\left[1+b_1(\Omega_{b0}h^2)^{b2}\right]~,
\end{equation}
where
\begin{eqnarray}\label{Defb1b2}
b_1 &=& 0.313(\Omega_{m0}h^2)^{-0.419}\left[1+0.607(\Omega_{m0}h^2)^{0.674}\right], \\
b_2 &=& 0.238(\Omega_{m0}h^2)^{0.223}.
\end{eqnarray}

For the SDSS DR7 data \cite{SDSSDR7}, we write $\chi^2$ for the BAO
data as
\begin{equation}
\chi^2_{BAO,SDSS}=\Delta{p_i}(C_{BAO,SDSS}^{-1})_{ij}\Delta p_j,
\end{equation}
where
\begin{equation}
\Delta p_i = p^{\rm data}_i - p_i,
\ \  p^{\rm data}_1 = d^{\rm data}_{0.2 } = 0.1905,
\ \  p^{\rm data}_2 = d^{\rm data}_{0.35} = 0.1097,
\end{equation}
and the inverse covariance matrix takes the form
\begin{equation}
(C_{BAO,SDSS}^{-1})=\left(
  \begin{array}{cc}
    30124  & -17227 \\
    -17227 & 86977 \\
  \end{array}
\right).
\end{equation}

The 6dFGS survey \cite{6dFGS} gives a measurement of $d_{0.106}^{\rm
data}=0.336\pm0.015$, so we have
\begin{equation}
\chi^2_{BAO,6dFGS}=\left(\frac{d_{0.106}-0.336}{0.015} \right)^2.
\end{equation}

The WiggleZ Dark Energy Survey \cite{WiggleZ} gives three measurements of
the $A$ parameter in different redshifts. The $A$ parameter is
defined by \cite{Eisenstein}
\begin{equation}
A(z)\equiv \frac{100 D_V(z)\sqrt{\Omega_{m0}h^2}}{z}.
\end{equation}
and the $\chi^2$ of the WiggleZ Dark Energy Survey takes the form
\begin{equation}
\chi^2_{BAO,WiggleZ}=\Delta{p_i}(C_{BAO,WiggleZ}^{-1})_{ij}\Delta p_j,
\end{equation}
where
\begin{equation}
\Delta p_i = p^{\rm data}_i - p_i,
\ \  p^{\rm data}_1 = A^{\rm data}_{0.44} = 0.474,
\ \  p^{\rm data}_2 = A^{\rm data}_{0.6} = 0.442,
\ \  p^{\rm data}_3 = A^{\rm data}_{0.73} = 0.424,
\end{equation}
and the inverse covariance matrix takes the form
\begin{equation}
(C_{BAO,WiggleZ}^{-1})=\left(
  \begin{array}{ccc}
    1040.3 & -807.5 & 336.8 \\
    -807.5 & 3720.3 & -1551.9 \\
    336.8 & -1551.9  &  2914.9 \\
  \end{array}
\right).
\end{equation}

The final BAO $\chi^2$ is a combination of the SDSS DR7, the 6dFGS
and the WiggleZ BAO $\chi^2$, i.e.,
\begin{equation}
\chi^2_{BAO} = \chi^2_{BAO,SDSS} + \chi^2_{BAO,6dFGS} + \chi^2_{BAO, WiggleZ}.
\end{equation}

\subsection{The Hubble constant data}

The precise measurements of $H_0$ will be helpful to break the
degeneracy between it and dark energy parameters \cite{H0Freedman}.
When combined with the CMB measurement, it can lead to precise
measure of the dark energy equation of state (EOS), $w$ \cite{H0WHu}. Recently, using
the WFC3 on the HST, Riess {\it et al.} obtained an accurate
determination of the Hubble constant \cite{HSTWFC3},
\begin{equation}
H_0=73.8\pm 2.4~ {\rm km/s/Mpc},
\end{equation}
corresponding to a $3.3\%$ uncertainty. So the $\chi^2$ of the Hubble
constant measurement is
\begin{equation}
\chi^2_{h}=\left({h-0.738\over 0.024}\right)^2.
\end{equation}

\subsection{The total $\chi^2$}

Since the SNIa, CMB, BAO and $H_0$ are effectively independent measurements,
we can combine them by simply adding together the $\chi^2$ functions,
i.e.,
\begin{equation}
\chi^2_{total} = \chi^2_{SN} + \chi^2_{CMB} + \chi^2_{BAO} +
\chi^2_{h}.
\end{equation}

\section{Results}

In this section, firstly, we show the cosmological constraints on the three IHDE models,
and then we discuss the cosmological implications of these models.

\subsection{Cosmological constraints}\label{Sec:constrains}

In this subsection, we discuss the cosmological constraints on the
three IHDE models. The parameter space of these IHDE models are
explored by using the Union2.1+BAO+CMB+$H_0$ data. We summarize the
fitting results in Table~\ref{table1}. For comparison, the fitting
results of the $\Lambda$CDM and HDE models by using the same set
of data are also shown.

The values of $\chi^2_{min}$ of the models are listed in column 7.
Clearly, compared with the $\Lambda$CDM and HDE models, the IHDE
models lead to a evident reduction of $\chi^2_{min}$. However,
taking the number of parameters into account, $\Lambda$CDM and HDE
models still provide nice fits to the data. In addition, compared
with the two simple IHDE models, the complicated model with
$Q\propto\rho_{dm}^\alpha\rho_{de}^\beta$ does not lead to a
significant reduction of $\chi^2_{min}$, implying that in the
context of the current observations, such a complicated model is not
necessary.

\begin{table} \caption{Fitting results of the models}
\begin{center}
\label{table1}
\begin{tabular}{lccccccc}
  \hline\hline
Model   &           $\Omega_{dm0}$      &             $c$                       &        $c$ (95.4\% range)      &               $\Gamma$         &~~             $\Omega_{k0}$     ~~&~~   $\chi^2_{min}$    \\
  \hline
  $\Lambda$CDM~~~~~~~~~~~~~~~~~~~~~~~~~
      &~~ $0.235^{+0.012}_{-0.011}$ ~~&~~        --         ~~ &~~         --         ~~ & ~~ 0 (fixed)         ~~&   $0.001^{+0.005}_{-0.005}$         &~~     550.354    ~~~  \\
  \hline
  HDE ($Q=0$)~~~~~~~~~~~~~~~
      &~~ $0.233^{+0.012}_{-0.011}$ ~~&~~ $0.71^{+0.10}_{-0.08}$~~ &~~ $0.56 \leq c \leq 0.94$~~ & ~~                            0 (fixed)~~&~ $0.006^{+0.007}_{-0.007}$ ~~&~~     549.461    ~~~   \\
  \hline
  $Q=\Gamma H_0 \rho_{dm}$~~~~~~~~~~~~~~~~~
      &~~ $0.238^{+0.013}_{-0.012}$ ~~&~~ $0.69^{+0.10}_{-0.08}$~~ &~~ $0.54 \leq c \leq 0.91$~~ & ~~$-0.056^{+0.051}_{-0.051}$~~&~~$0.015^{+0.011}_{-0.011}$ ~~&~~    548.352    ~~~  \\
  \hline
  $Q=\Gamma H_0 \rho_{de}$~~~~~~~~~~~~~~~~~~
      &~~ $0.237^{+0.013}_{-0.011}$ ~~&~~ $0.66^{+0.11}_{-0.09}$~~ &~~ $0.50 \leq c \leq 0.90$~~ & ~~$-0.073^{+0.070}_{-0.072}$~~&~~$0.016^{+0.012}_{-0.012}$ ~~&~~    548.390    ~~~\\
  \hline
  $Q=\Gamma H_0 \rho_{dm}^\alpha\rho_{de}^\beta/\rho_{c0}^{\alpha+\beta-1}$
      &~~ $0.239^{+0.013}_{-0.013}$ ~~&~~ $0.63^{+0.18}_{-0.08}$~~ &~~ $0.47 \leq c \leq 0.92$~~ & ~~$-0.014^{+0.014}_{-0.237}$~~&~~ $0.017^{+0.011}_{-0.014}$ ~~&~~   548.298    ~~~ \\
 \hline\hline
\end{tabular}
\end{center}
\end{table}

In column 2--6, the best-fit values and the 68.3\% CL ($\Delta
\chi^2=1$) uncertainties of the parameters $\Omega_{dm0}$, $c$,
$\Gamma$ and $\Omega_{k0}$ are listed. For the HDE and three IHDE
models, we found $c < 1$ at the 95.4\% CL ($\Delta \chi^2=4$). This
result is consistent with the results in the previous works
\cite{holoobs09,intholo}. For the HDE model, $c<1$ means that the
universe will end up with a the big rip, while for the IHDE models,
due to the interaction between the dark sectors, the big rip may be
avoided. We will discuss this issue in the following content.

For these three IHDE models, we find $\Gamma\lesssim0$ at the 68.3\%
CL, namely, the energy flow from dark matter to dark energy is
slightly favored by the data.

\begin{figure}
\includegraphics[height=7cm]{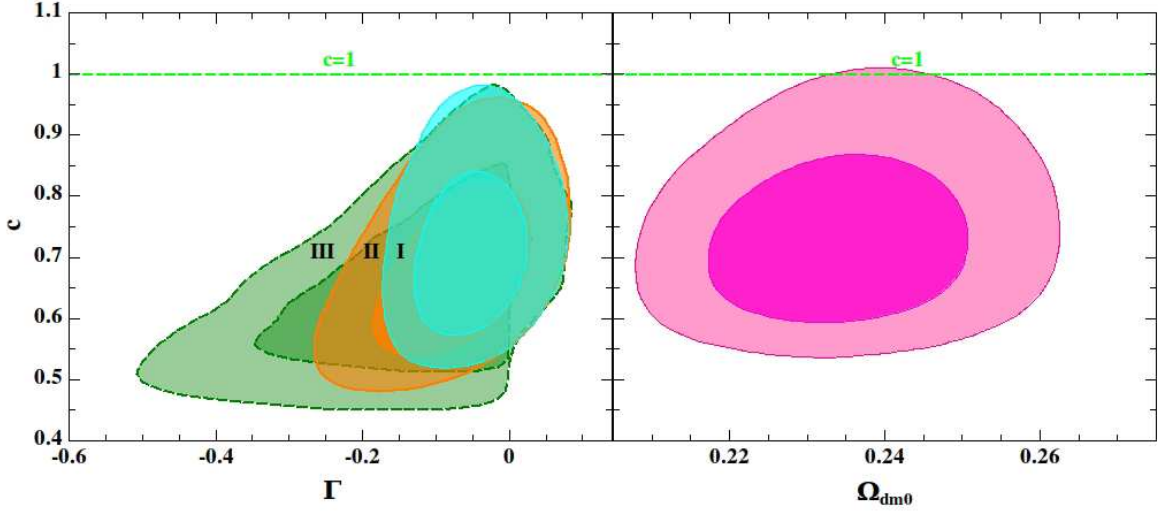}
\caption{\label{Fig:c} Marginalized probability contours at the
$68.3\%$ ($\Delta \chi^2 = 2.3$) and $95.4\%$ ($\Delta \chi^2 =
6.18$) CLs in the $\Gamma$--$c$ (for the three IHDE models) and
$\Omega_{dm0}$--$c$ (for the HDE model) planes. The contours for the
IHDE1 (cygan, labeled as I), IHDE2 (orange, labeled as II), and
IHDE3 (olive, labeled as III) models are plotted in the left panel,
and the contours for the HDE model (magnetic) are plotted in the
right panel. The green dashed line denotes $c=1$. Clearly, compared
with the HDE model, the fitting results of $c$ in the three IHDE
models are smaller.}
\end{figure}

In the left panel of Fig.~\ref{Fig:c}, we plot the contours of the
68.3\% and 95.4\% CL for the three IHDE models in the $\Gamma$--$c$
plane. We can see that the upper bounds on the parameter $c$ are
similar, while the lower constraints are slightly different from
each other. In the IHDE3 model, due to the complexity of the model,
the allowed parameter space is much larger than those of the IHDE1
and IHDE2 models.

Interestingly, it seems that, compared with the HDE model, the IHDE
models slightly favor smaller values of $c$. This can be seen in the
column 4 of Table~\ref{table1}. To see it more clearly, we also plot
the probability contours in the $\Omega_{dm0}$--$c$ plane for the
HDE model in the right panel of Fig.~\ref{Fig:c}. The 95.4\% CL of
the HDE model intersects with the line $c=1$, while the contours of
the IHDE models are all below the line.

\begin{figure}
\includegraphics[height=7cm]{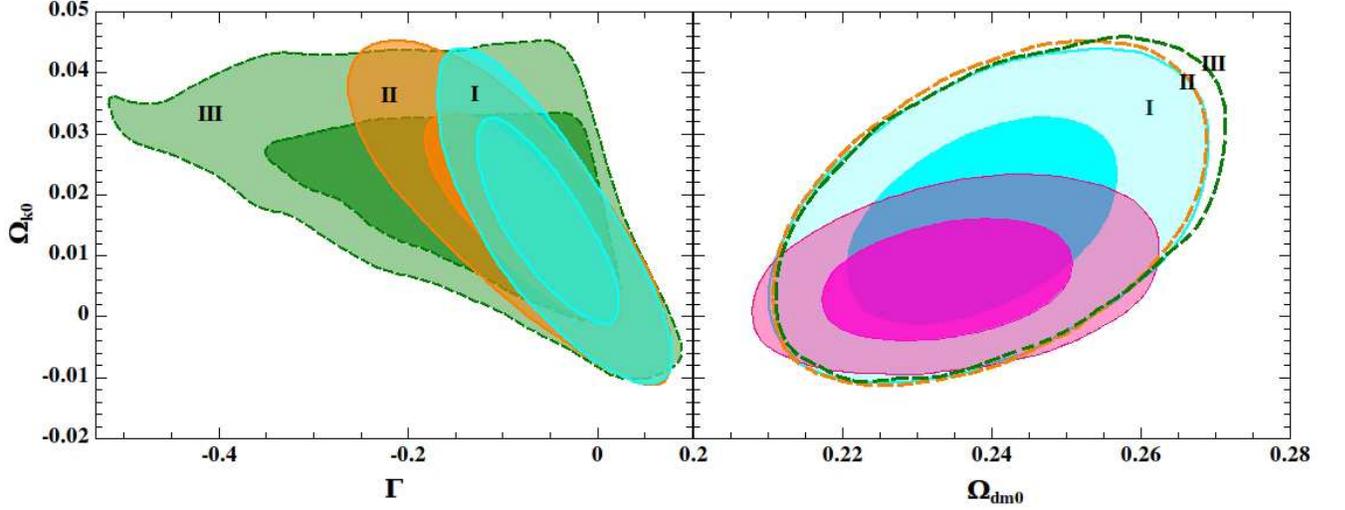}
\caption{\label{Fig:Omegak0} Marginalized probability contours at
the $68.3\%$ and $95.4\%$ CLs in the $\Gamma$--$\Omega_{k0}$ (left
panel) and $\Omega_{dm0}$--$\Omega_{k0}$ (right panel) planes for
the IHDE1 (cygan, labeled as I), IHDE2 (orange, labeled as II),
IHDE3 (olive, labeled as III), and the HDE (magnetic) models. The
$\Omega_{dm0}$--$\Omega_{k0}$ contours of the three IHDE models are
similar to each other, so for the IHDE2 and IHDE3 models, we only
plot the 95.4\% CL contours.}
\end{figure}

In the left panel of Fig.~\ref{Fig:Omegak0}, we plot the contours of
the 68.3\% and 95.4\% CL for the three IHDE models in the
$\Gamma$--$\Omega_{k0}$ plane. The parameter space of the IHDE3
model is much larger than those of the IHDE1 and IHDE2 models, and
we can see clear degeneracy between $\Omega_{k0}$ and $\Gamma$,
consistent with the result of \cite{KIHDE}. The degeneracy amplifies
the range of $\Omega_{k0}$ compared with the HDE model without
interaction, as is seen in the right panel of
Fig.~\ref{Fig:Omegak0}, where the contours in the
$\Omega_{dm0}$--$\Omega_{k0}$ plane for the three IHDE and the HDE
models are all plotted.

\begin{figure}
\includegraphics[height=7cm]{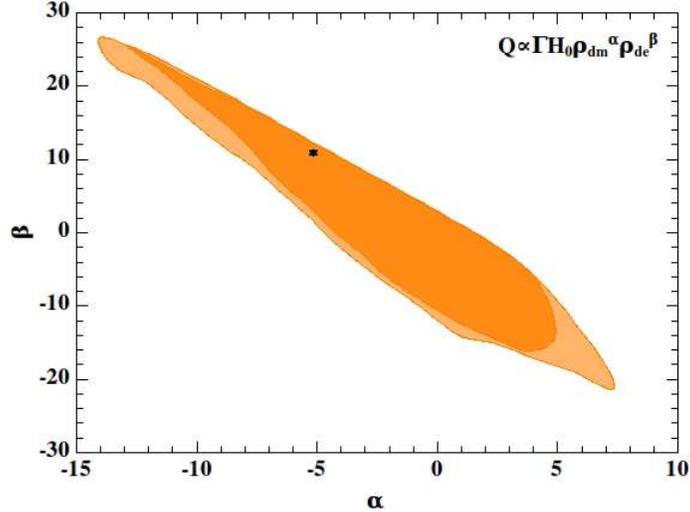}
\caption{\label{Fig:alphabeta} Marginalized probability contours at
the $68.3\%$ and $95.4\%$ CLs in the $\alpha$--$\beta$ planes for
the IHDE3 model.}
\end{figure}

We are also interested in the constraints on the parameters $\alpha$
and $\beta$, i.e., the interaction forms allowed by the data. In our
analysis, we find the 95.4\% CL constraints are
\begin{equation} \label{eq:alphabeta}
\alpha = -5.17^{+13.37}_{-9.59},\ \ \ \beta = 10.98^{+18.75}_{-33.42}.
\end{equation}
So the allowed regions of $\alpha$ and $\beta$ are very wide. This
phenomenon is understandable: the intensity of the interaction term
$Q$ is almost decided by $\Gamma$. If $\Gamma$ is small enough, any
large values of $\alpha$ and $\beta$ are tolerable. \footnote{In our
analysis we generate about $10^7$ samples for the IHDE3 model. It is
expected that the area of the contour could be larger if one
performs a more extensive analysis.} The 68.3\% and 95.4\% CL
contours in the $\alpha$--$\beta$ plane are plotted in
Fig.~\ref{Fig:alphabeta}, and we can see strong degeneracy between
$\alpha$ and $\beta$.

\subsection{Constraints on the interaction term}\label{sec:interaccons}

In this subsection we discuss the constraints on the interaction
term in the IHDE models. The fitting results of $\Gamma$ have been
listed in the 5-{\it th} column of Table~\ref{table1}. We have seen
that for all the models we have $\Gamma \leq 0$ at the 68.3\% CL.
Here we list the 95.4\% CL constraints for the three IHDE models,
\begin{eqnarray}
-0.154 &\leq& \Gamma \leq 0.053\ \ \ \ \rm for \ IHDE1,\\
-0.229 &\leq& \Gamma \leq 0.059\ \ \ \ \rm for \ IHDE2,\\
-0.509 &\leq& \Gamma \leq 0.072\ \ \ \ \rm for \ IHDE3.
\end{eqnarray}
For the $\Gamma>0$ case, i.e., the direction of the energy flow is
from dark energy to dark matter, we find a tight constraint
$|\Gamma|\lesssim 0.07$ at the 95.4\% CL for all the three IHDE
models. For the $\Gamma<0$ case, the allowed value of $\Gamma$ is
much larger. At the 95.4\% CL, for the IHDE1 and IHDE2 models we
have $|\Gamma|\lesssim 0.2$, while for the IHDE3 model we have
$|\Gamma|\lesssim 0.5$.
\begin{figure}
\includegraphics[height=7cm]{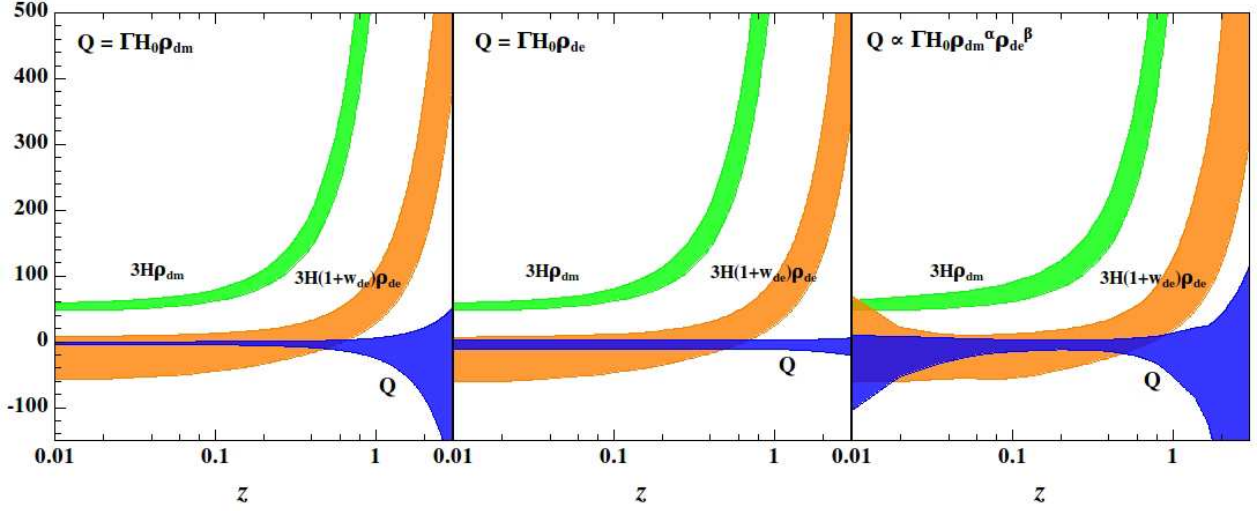}
\caption{\label{Fig:rhoQs}Reconstructed evolutions of $3H\rho_{dm}$,
$3H(1+w_{eff})\rho_{de}$ and $Q$ (all divided by $\rho_{c0}$) at the
95.4\% CL, for the IHDE1 (left panel), IHDE2 (middle panel) and
IHDE3 (right panel) models, respectively.}
\end{figure}

To see the result more vividly, in Fig.~\ref{Fig:rhoQs} we
reconstruct the evolution of $3H\rho_{dm}$, $3H(1+w_{de})\rho_{de}$,
and $Q$ at the 95.4\% CL for the three IHDE models. As shown in
Eqs.~(\ref{eq:IHDEeq0}) and (\ref{eq:IHDEeq1}), the term
$3H\rho_{dm}$ [or, $3H(1+w_{de})\rho_{de}$] characterizes the
influence of the cosmic expansion on the energy density of dark
matter (or, dark energy), while $Q$ describes the effect of the
interaction. As shown in Fig. \ref{Fig:rhoQs}, for all the models
$Q$ is much smaller compared with $3H\rho_{dm}$ and
$3H(1+w_{de})\rho_{de}$, implying that the influence of interaction
on $\dot\rho_{dm}$ and $\dot\rho_{de}$ is much weaker than the
cosmic expansion. It is also interesting to find that the evolution
of the interaction terms have different behaviors in the three IHDE
models, due to the different expressions of $Q$. In the IHDE1 model with
$Q \propto \rho_{dm}$, the absolute value of $Q$ may significantly
decrease (along with $\rho_{dm}$) during the evolution, while in the
IHDE2 model the absolute value of $Q$ almost maintains no changes
during the evolution since $Q\propto \rho_{de}$.

We are interested in the influence of the interaction on the dark
matter density, and thus we investigate the quantity
$Q/(3H\rho_{dm})$, which describes the ratio between the changes of
$\rho_{dm}$ caused by the interaction and by the cosmic expansion.
Here we list the constraints on this quantity at $z=0$ at the 95.4\%
CL,
\begin{eqnarray}\label{Qratioz0}
-0.0514 \leq \frac{Q}{3H\rho_{dm}} \leq 0.0177\ \ \ \ \rm for \ IHDE1,\\
-0.2097 \leq \frac{Q}{3H\rho_{dm}} \leq 0.0625\ \ \ \ \rm for \ IHDE2,\\
-0.4779 \leq \frac{Q}{3H\rho_{dm}} \leq 0.0105\ \ \ \ \rm for \ IHDE3.
\end{eqnarray}
We see that in the IHDE1 model, the present effect of $Q$ is much
smaller than the other two IHDE models. The large value (47.79\%) of
the lower bound of this ratio in the IHDE3 model indicates that in
this model the influence of $Q$ on $\dot\rho_{dm}$ can be comparable
with that of the cosmic expansion. Figure~\ref{Fig:rhoQs} shows that
in the IHDE1 model the present value of $Q$ is much tightly
constrained compared with the other two models, and in the IHDE3
model $Q$ could take large negative values when $z\rightarrow0$.

Also, since we are interested in {\it how much dark matter is
produced or annihilated} due to the interaction, it is helpful to
define another quantity,
\begin{equation} \label{eq:Delta_def}
\Delta \equiv
\frac{\rho_{dm}a^3\big|_{z=0}-\rho_{dm}a^3\big|_{ini}}{\rho_{dm}a^3\big|_{ini}}.
\end{equation}
Notice that $\rho_{dm} a^3$ is the total energy of dark matter in a
unit comoving volume. In the absence of interaction, $\rho_{dm} a^3$
is conserved, and $\Delta=0$. Here $\rho_{dm}a^3\big|_{ini}$ is the
``initial'' value of the total energy of dark matter in a comoving
volumn. It can be calculated at the early epoch, e.g., $z>5000$.
Thus, $\Delta$ describes the relative change amount of energy in the
dark matter component caused by the interaction.

The results for the three IHDE models are as follows:
\begin{eqnarray}\label{eq:Delta}
-15.22\% &\leq& \Delta \leq 8.83\%\ \ \ \ \rm for \ IHDE1,\\
-13.3\% &\leq& \Delta \leq 5.57\%\ \ \ \ \rm for \ IHDE2,\\
-14.49\% &\leq& \Delta \leq 6.57\%\ \ \ \ \rm for \ IHDE3.
\end{eqnarray}
We find that the results of the three models are not much different
from each other. Thus, although the evolution of the interaction
term is evidently different, the constraint on the overall effect of
the interaction does not change a lot. Roughly, at the 95.4\% CL,
the increment amount of the energy of dark matter is constrained to
be less than 9\%, while the decrement amount is constrained to be
less than 15\%.

\subsection{The fate of the universe}

In this subsection we discuss the fate of the universe in the IHDE
models. The most important reason for the introduction of the
interaction between the dark sectors in the HDE model is to avoid
the future big-rip singularity. For the IHDE model with $Q\propto
\rho_{dm}^\alpha\rho_{dm}^\beta$, this issue has been briefly
discussed in \cite{YZMa}, where it is shown that back reaction
effect from quantum corrections cannot prohibit the occurrence of
the big rip. Here, we will give a more detailed investigation of the
fate of the universe in the IHDE models considered above.

Firstly, we will derive some useful formulas as a preparation of our
discussion. Then, we will analyze the dynamical equations to see
whether it is possible to avoid the big rip when $c<1$. Finally,
based on the constraints of the data, we will numerically solve the
equations and reconstruct the evolution of the dark sectors to
$z\rightarrow-1$.

\subsubsection{The effective EOS of dark sectors}

We can define the effective pressure and effective EOS of dark energy, i.e.,
\begin{equation}\label{eq:weffdef}
 p_{eff,de} = p_{de}+\frac{Q}{3H},\ \ \ w_{eff,de}=w_{de}+\frac{Q}{3H\rho_{de}},
\end{equation}
satisfying
\begin{equation}
 \dot\rho_{de}+3H(\rho_{de}+p_{eff})=0,\ \ \ \dot\rho_{de}+3H(1+w_{eff,de})\rho_{de}=0.
\end{equation}

From Eq.~(\ref{eq:pde}), using $\frac{d}{dt}=-(1+z)H\frac{d}{dz}$
and $\frac{\dot H}{H^2}=-(1+z)\frac{d\ln H}{dz}=-(1+z)\frac{d\ln
E}{dz}$, we derive the equation of state for dark energy,
\begin{eqnarray}
 w_{de} = \frac{2(1+z)}{3\Omega_{de}}\frac{d\ln E}{dz}-\frac{1}{\Omega_{de}}
        -{1\over3}\frac{\Omega_r}{\Omega_{de}}+{1\over3}\frac{\Omega_k}{{\Omega_{de}}}.
\end{eqnarray}
Combined with Eq.~(\ref{eq:OH3}), it follows that
\begin{eqnarray}
w_{de} = -{1\over3}-{2\over3}\sqrt{{\Omega_{de}\over c^2}+\Omega_k}-\frac{\Omega_I}{3\Omega_{de}},
\end{eqnarray}
which reduces to the familiar formula
$w_{de}=-\frac{1}{3}-\frac{2}{3}\frac{\sqrt{\Omega_{de}}}{c}$ when
$\Omega_k=0$ and $\Omega_I=0$. Thus, the effective EOS takes the
form
\begin{eqnarray}
 w_{eff,de} &=& w_{de} + \frac{Q}{3H\rho_{de}} \label{eq:weffde1}\\
         &=& -{1\over3}-{2\over3}\sqrt{{\Omega_{de}\over c^2}+\Omega_k}.\label{eq:weffde2}
\end{eqnarray}
This result is very interesting: $w_{eff,de}$ does not explicitly
contain the interaction term. However, we should keep in mind that
the evolutions of $\Omega_{de}(z)$ and $E(z)$ are determined by the
differential equations (\ref{eq:OH3}) and (\ref{eq:OH4}) where the
interaction $Q$ (or, $\Omega_I$) is involved.

It is much easier to derive the effective EOS of dark matter. From
Eq.~(\ref{eq:IHDEeq1}), it follows directly that
\begin{eqnarray}
 w_{eff,dm}  &=& -\frac{Q}{3H\rho_{dm}} = \frac{\Omega_I}{3\Omega_{dm}} \\
             &=& -\frac{1}{3}\Gamma E(z)^{2(\alpha+\beta)-3}\Omega_{dm}^{\alpha-1}\Omega_{de}^\beta.
\end{eqnarray}

To investigate the fate of the universe, we are also interested in
the effective EOS of the total components in the universe, i.e.,
\begin{equation}
 w_{eff,tot} = \sum_i w_{eff,i} \Omega_i,
\end{equation}
where $i$ represents $de$, $dm$, $b$, $r$ and $k$. In the future
epoch, only the dark energy and dark matter components are
important.

\subsubsection{Analysis of the dynamical equations in the $z\rightarrow-1$ region}

Now we make some general investigations to see whether the big rip
can be avoided when $c<1$ in the IHDE models.
Equation~(\ref{eq:weffde2}) shows that the effective EOS of HDE is
only determined by $\Omega_{de}$ and $c$ (we neglect $\Omega_{k}$
which is less important in the future). In the case of $c<1$, once
$\Omega_{de} \rightarrow 1$ there will always be $w_{eff,de}<-1$ and
$w_{eff,tot}<-1$, and the fully dominated dark energy will drive the
universe to a big rip. Thus, to avoid the big rip, we must evade
$\Omega_{de}\rightarrow1$ in the future. Also, to avoid $\rho_{de}$
from infinitely increasing, we must require $\Gamma > 0$, i.e., the
direction of energy flow is from dark energy to dark matter.

Here we list the simplified equations of motion in the region
$z\rightarrow-1$. Neglecting the radiation, baryon and curvature
components, we have
\begin{equation}
\label{simeq:OH3}{1\over E(z)}{dE(z) \over dz}
=-{\Omega_{de}\over
1+z}\left({\Omega_{de}-3+\Omega_I\over2\Omega_{de}}+\frac{\sqrt{\Omega_{de}}}{c} \right),
\end{equation}
\begin{equation}
\label{simeq:OH4}
{d\Omega_{de}\over dz}=
-{2\Omega_{de}(1-\Omega_{de})\over 1+z}\left(\frac{\sqrt{\Omega_{de}}}{c}+\frac{1}{2}-{\Omega_I\over 2(1-\Omega_{de})}\right).
\end{equation}
Notice that unlike the cases of $Q\propto H\rho_{dm}$ or $Q\propto
H\rho_{de}$, the two equations are coupled, and thus the situation
is much more complicated. However, we find that these equations have
a stable point
\begin{equation}
 \Omega_{de} = c^2, \ \ \Omega_{I}=3(1-c^2),
\end{equation}
which leads to
\begin{equation}
 w_{eff,de}=-\frac{1}{3}-\frac{2\sqrt{\Omega_{de}}}{3c}=-1, \ \ \ w_{eff,dm}=-\frac{\Omega_I}{3\Omega_{dm}}=-1,\ \ \
\end{equation}
and
\begin{equation}
 \frac{dE(z)}{dz}=0,\ \ \ \frac{d\Omega_{de}}{dz}=0.
\end{equation}
This is exactly a de Sitter solution. See also \cite{XZ10} for a
relevant study about the avoidance of big rip in the holographic
dark energy model, where the extra dimension effect is considered
and the final state of the universe is also a de Sitter solution.

\begin{figure}
\includegraphics[height=8cm]{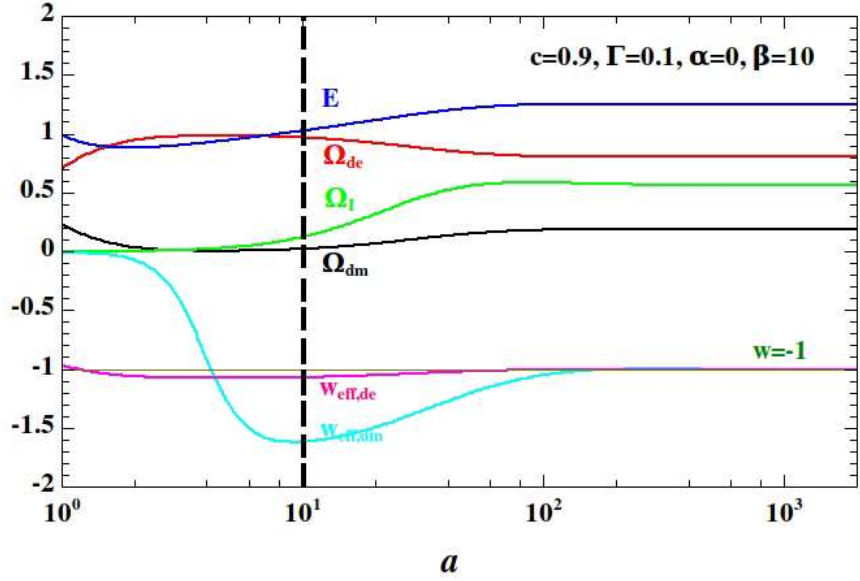}
\caption{\label{Fig:Demo} Evolutions of $E$, $\Omega_{de}$,
$\Omega_{dm}$, $\Omega_{I}$, $w_{eff,de}$ and $w_{eff,dm}$ along
with the scale factor $a$. Here we take typical values of
$\Omega_{dm0}=0.24$, $\Omega_{k0}=0$, $h=0.73$, and others are
denoted in the panel. }
\end{figure}

We find that this situation can really happen when $\beta$ is large,
and the interaction is strong enough to pull down $\Omega_{de}$ from
1. As an example, in Fig.~\ref{Fig:Demo} we demonstrate how the
quantities evolve along with the scale factor $a$ in the case of
$\beta=10$. A sophisticated discussion on the evolution behaviors of
the quantities in the whole range of $a$ would be tedious and
unnecessary. Thus, instead, we just focus on the epoch of $a>10$
(the region to the right of the thick dashed line). This would be
enough for us to understand how it happens. In this epoch, firstly,
$E(z)$ increases due to the dominated dark energy component with
$w_{eff,de}<-1$, while the strong interaction pulls $\Omega_{de}$
down from 1. So along with $a$, $\Omega_{de}$ decreases and
correspondingly $\Omega_{dm}$ increases. As a result, $w_{eff,de}$
increases from values less than $-1$, while the effective EOS of
dark matter, $w_{eff,dm}=\frac{\Omega_I}{3\Omega_{dm}}$, which is
highly negative due to a small $\Omega_{dm}$, also goes up. When
both $w_{de,eff}$ and $w_{dm,eff}$ approach $-1$, the system reaches
the stable point, and $E(z)$ stops increasing. The final state of
the universe is de Sitter-like.

\begin{figure}
\includegraphics[height=5.5cm]{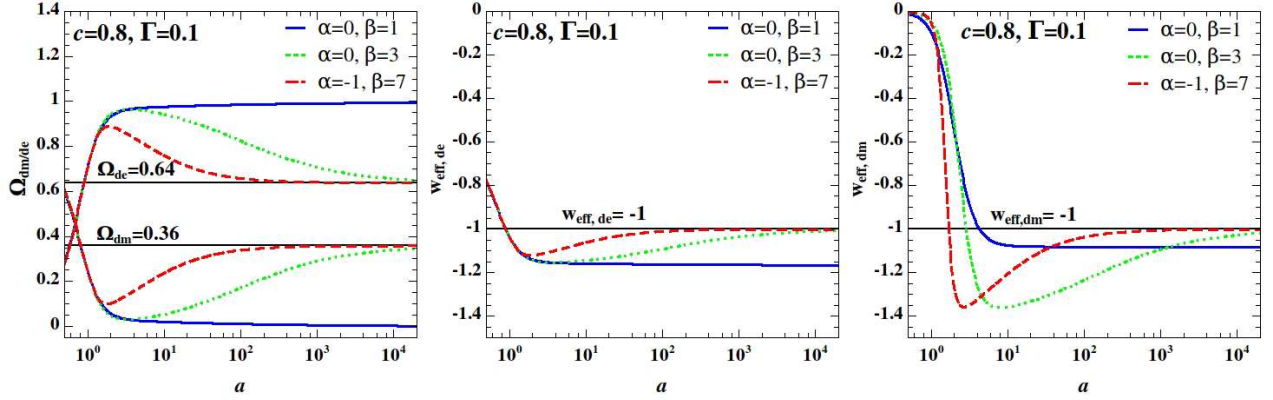}
\caption{\label{Fig:Omegaweffs} Evolutions of $\Omega_{dm}$ (left
panel), $\Omega_{de}$ (left panel), $w_{eff,de}$ (middle panel) and
$w_{eff,dm}$ (right panel) along with the scale factor $a$ for three
sets of $\alpha$ and $\beta$. The same as Fig.~\ref{Fig:Demo}, we
take values of $\Omega_{dm0}=0.24$, $\Omega_{k0}=0$ and $h=0.73$.}
\end{figure}

Note that an essential condition that such a procedure can happen is
that the interaction is strong enough to pull $\Omega_{de}$ down
from 1. Thus, a large $\beta$ is expected. To see the dependence on
the values of $\beta$, in Fig. \ref{Fig:Omegaweffs} we plot the
evolution of $\Omega_{de}$, $\Omega_{dm}$, $w_{eff,de}$ and
$w_{eff,dm}$ in large $a$ regions for different values of $\alpha$
and $\beta$. For the case of $\alpha=0$ and $\beta=1$ (the blue
line), the interaction is not strong enough to avoid
$\Omega_{de}\rightarrow 1$, so the big rip still happens. For the
case of $\alpha=0$ and $\beta=3$ (the green dotted line), the
interaction is strong enough, and so we see that $\Omega_{de}$ is
pulled down from 1, and the system evolves to a stable point when
$a>10000$. For the case of $\alpha=-1$ and $\beta=7$ (the red dashed
line), the interaction is so strong that the system quickly
approaches the stable point at $a< 1000$.

In the following, we will numerically solve the equations for the
IHDE models in a large $a$ region in the parameter space constrained
by the data, and see whether the de Sitter solution can be achieved
in these models.

\subsubsection{The IHDE1 and IHDE2 models}

\begin{figure}
\includegraphics[height=10cm]{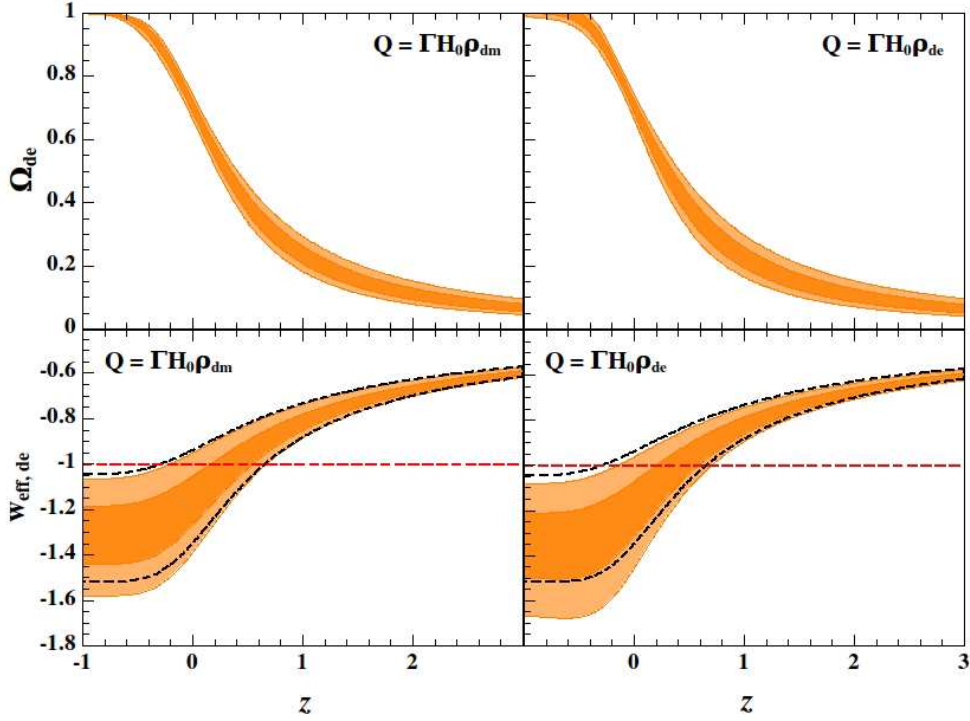}
\caption{\label{Fig:weffOde} The evolutions of $\Omega_{de}$ (top
panels) and $w_{eff,de}$ (bottom panels) along with the redshift $z$
at the 68.3\% and 95.4\% CLs, for the IHDE1 (left panels) and IHDE2
(right panels) models. The 95.4\% CL of HDE model is plotted in
dashed lines for a comparison.}
\end{figure}

In the top panels of Fig. \ref{Fig:weffOde} we show the
reconstructed evolution of $\Omega_{de}$ along with $z$ for the
IHDE1 and IHDE2 models. Clearly, at the 95.4\% CL we see
$\Omega_{de}\rightarrow1$ when $z\rightarrow-1$, and we have
$w_{eff,de}<-1$ since $c<1$. So these models will not help us to
avoid the big rip.

This result is understandable. The fate of the universe in the IHDE
model with the interaction term
\begin{equation}
 Q = 3bH\rho_{de}
\end{equation}
was discussed in \cite{HDEstable} and it was shown that the
condition to prevent the big rip is
\begin{equation}\label{eq:bigripcond}
 b \geq c^{-2}-1.
\end{equation}
In the IHDE1 and IHDE2 models, analogously, we define the
``effective'' coupling,
\begin{equation}
 b_{eff} \equiv \frac{Q}{3H\rho_{de}},
\end{equation}
and we have
\begin{equation}
 b_{eff} = \frac{\Gamma\Omega_{dm}}{3E(z)\Omega_{de}},~~~{\rm and}~~~ b_{eff}=\frac{\Gamma}{3E(z)}
\end{equation}
for the two models, respectively. Note that once $E(z)$ becomes
increasing in the future, $b_{eff}$ will be suppressed to zero, so
the condition (\ref{eq:bigripcond}) cannot be satisfied.

The reconstructed $w_{eff,de}(z)$ for the two models are shown in
the lower panels of Fig. \ref{Fig:weffOde}. As a comparison, the
95.4\% CL evolution of the HDE model is also plotted (the black
dashed line). We see that for these models
$w_{eff,de}(z\rightarrow-1)<-1$, and thus the big rip will happen.
Also, the values of $w_{eff,de}$ in the IHDE models are more
negative than that of the HDE model, since in the two IHDE models
$c$ is constrained to be smaller values. Thus, instead of helping us
to avoid the big rip, the situation is even worse in these two
models.

\subsubsection{The IHDE3 model}

\begin{figure}
\includegraphics[height=7cm]{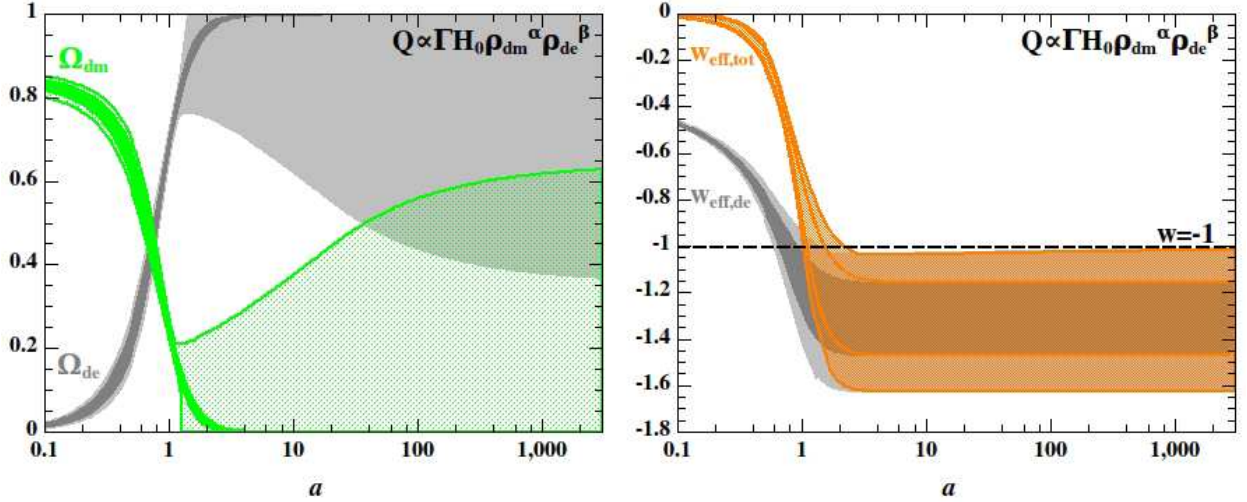}
\caption{\label{Fig:Omegaweff3} The evolutions of $\Omega_{de}$
(gray, left panel), $\Omega_{dm}$ (green, left panel), $w_{eff,de}$
(gray, right panel) and $w_{eff,tot}$ (orange, right panel) along
with the scale factor $a$ for the IHDE3 model. The 68.3\% and 95.4\%
CLs are plotted. }
\end{figure}

We have seen in Fig. \ref{Fig:Omegaweffs} that a large $\beta$ can drive the system to a stable point and avoid the big rip.
As shown in Eq. (\ref{eq:alphabeta}) and Fig. \ref{Fig:alphabeta},
in our fitting results we found that large values of $\beta$ are allowed,
so we expect that the big rip can be avoided in the IHDE3 model.

In Fig. \ref{Fig:Omegaweff3} we plot the evolutions of $\Omega_{de}$
(gray, left panel), $\Omega_{dm}$ (green, left panel), $w_{eff,de}$
(gray, right panel) and $w_{eff,tot}$ (orange, right panel) along
with the scale factor $a$ for the IHDE3 model. From the left panel,
we see that, at the 68.3\% CL $\Omega_{de}\rightarrow1$ is present
in the future due to a negative $\Gamma$ (see Table~\ref{table1}),
but the interaction can prevent $\Omega_{de}$ from approaching 1 at
the large $a$ region at the 95.4\% CL. The right panel shows that
the stable point (corresponds to $w_{eff,de}=w_{eff,tot}=-1$) can be
accomplished when $a > 1000$ at the 95.4\% CL. Thus, the IHDE3 model
may help to avoid the big rip singularity.

\begin{figure}
\includegraphics[height=14cm]{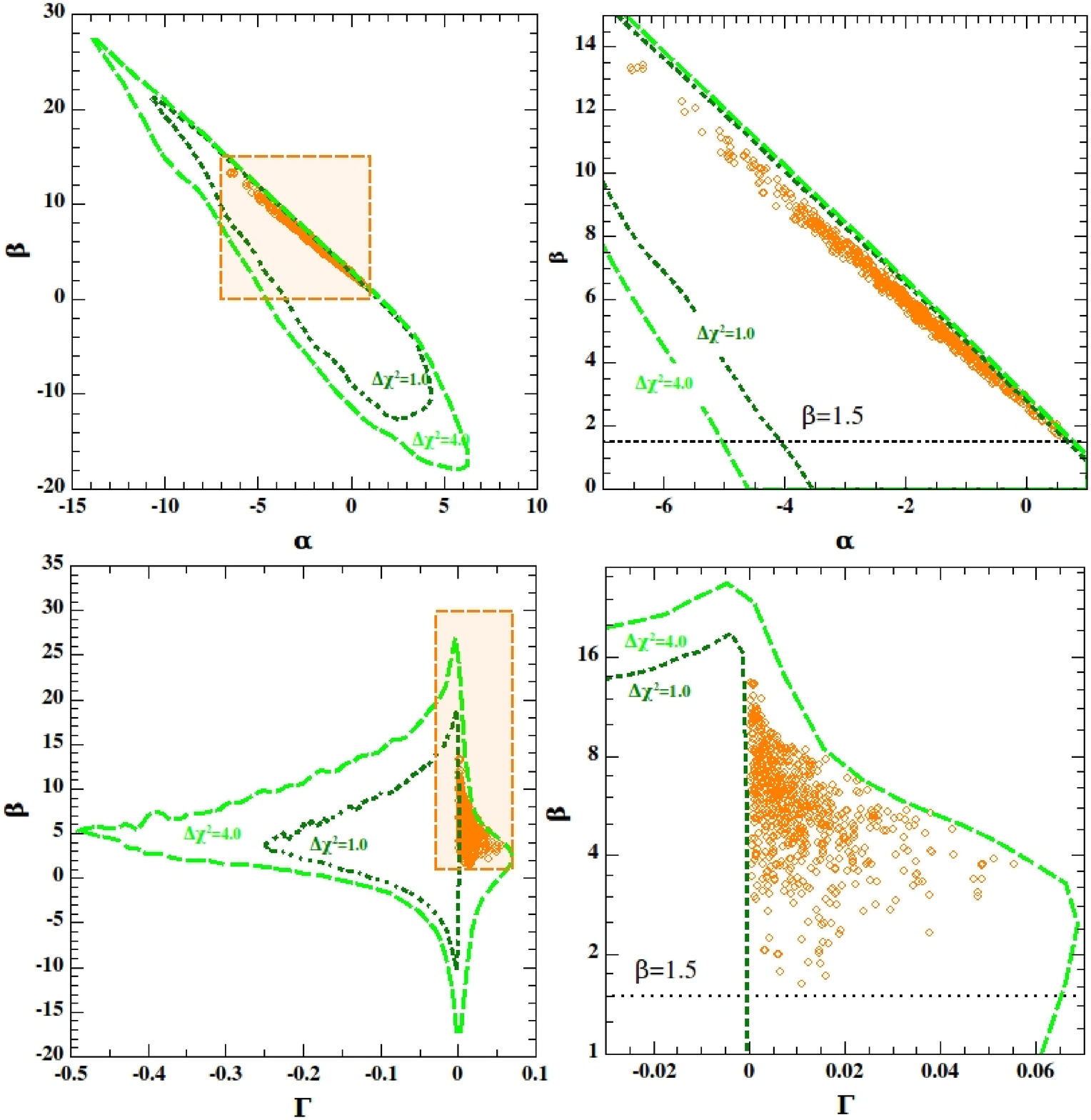}
\caption{\label{Fig:points} $\Delta \chi^2=1$ (olive dotted) and
$\Delta \chi^2=4$ (green dashed) contours in the $\alpha$--$\beta$
(top panels) and $\Gamma$--$\beta$ (bottom panels) planes. Orange
dots are the 709 samples which can achieve a de Sitter solution in
the future and thus avoid the big rip. In the left panels the whole
contours are plotted, while the right panels, as the zoom in of the
dashed orange region of the left panels, show these samples in
detail. $\beta=1.5$ is plotted in black dotted lines.}
\end{figure}

It would be worthy further investigating what kind of samples in our
MC analysis may achieve the de Sitter solution. Of $10^7$ samples
generated for the IHDE3 model there are about 4.7 million samples
satisfying $\Delta \chi^2<4$. Within them, we find about 18,000
samples satisfying $\Gamma>0$, and finally 709 samples having de
Sitter solution in the far future. In Fig. \ref{Fig:points}, these
samples are plotted in the $\alpha$--$\beta$ (top panels) and
$\Gamma$--$\beta$ (bottom panels) planes. Clearly, we see that to
achieve a de Sitter solution, a large $\beta$ is required (all these
samples have $\beta>1.5$). The bottom right panel of the figure
shows that, at the 68.3\% ($\Delta \chi^2<1$) CL, due to a negative
$\Gamma$, the big rip cannot be avoided.

\section{Concluding Remarks}

We have investigated the IHDE models with
$Q\propto\rho_{dm}^\alpha\rho_{de}^\beta$. Three IHDE models,
including the IHDE1 model with $Q=\Gamma H_0 \rho_{dm}$, the IHDE2
model with $Q=\Gamma H_0 \rho_{de}$, and the IHDE3 model with
$\alpha$ and $\beta$ running freely, were investigated.

By using the Union2.1+BAO+CMB+$H_0$ data, we placed cosmological
constraints on these models. We found that a negative $\Gamma$,
i.e., energy flow from dark matter to dark energy, is slightly
favored by the data, although the HDE model with $\Gamma=0$ still
lies in the 95.4\% CL region. For all the models, we get $c<1$ at
the 95.4\% CL. The values of $c$ in the IHDE models are smaller than
that in the HDE model.

We showed that the interaction has different properties in the three
models. Compared with the cosmic expansion, the effect of
interaction on the evolution of $\rho_{dm}$ and $\rho_{de}$ is
smaller. We also put a constraint on the total amount of the energy
change in the dark matter component. At the 95.4\% CL, the increment
amount of the dark matter is constrained to be less than 9\%, while
the decrement amount is constrained to be less than 15\%. We found
that the constraint basically does not depend on the forms of the
interaction term.

Furthermore, we discussed the fate of the universe by investigating
the dynamical equations of the IHDE models. We find that the
equations may give a de Sitter solution at $z\rightarrow-1$, with
the effective equation of state for dark energy and dark matter
being $-1$. When confronted with data, we show that this solution
cannot be accomplished in the IHDE1 and IHDE2 models. Rather, in
these two models since the data favor a smaller $c$, the big rip is
even more severe than the HDE model. In the IHDE3 model, we show
that such a solution can be achieved for a large $\beta$, and the
big rip may be avoided at the 95.4\% CL.

Finally, there are still some issues not covered in our paper, i.e.,
the {\it precise} condition for obtaining the de Sitter solution,
the coincidence problem, and so on. These issues all deserve further
investigations.

\begin{acknowledgments}
This work was supported by the NSFC under Grant Nos.~10535060,
10975172, 10821504, 10705041, 10975032 and 11175042, by the National
Ministry of Education of China under Grant Nos.~NCET-09-0276 and
N100505001, and by the 973 program (Grant No.~2007CB815401) of the
Ministry of Science and Technology of China.
\end{acknowledgments}


\end{document}